\documentclass[10pt]{article}

\newcommand{\phib}{{\vec\phi}}
\newcommand{\sigb}{{\vec\sigma}}
\newcommand{\veck}{\mathitbf k}
\newcommand{\vecx}{\mathitbf x}
\newcommand{\vecy}{\mathitbf y}
\newcommand{\veci}{\mathitbf i}

\newcommand{\e}{{\mathrm e}}
\renewcommand{\d}{{\mathrm d}}
\newcommand{\irm}{{\mathrm i}}
\newcommand{\tr}{{\mathrm{Tr}}}

\newcommand{\Dphi}{{\mathrm D}\vec\phi}
\newcommand{\Lt}{L_\tau}

\DeclareMathAlphabet{\mathitbf}{T1}{cmr}{bx}{it}

\usepackage{epsfig,amsfonts,amssymb,newlfont,rotate,float,array}

\begin{document}
\title{Hybrid Monte Carlo algorithm for the Double Exchange Model}
\author{J. L. Alonso$^1$, L. A. Fern\'andez$^2$, F. Guinea$^3$, \\
V. Laliena$^1$ and V. Mart{\'\i}n-Mayor$^4$\\
\small
$^1$ Dep. de F{\'\i}sica Te\'orica, Facultad de Ciencias,
Univ. de Zaragoza, 50009 Zaragoza, Spain.\\
\small $^2$ Dep. de F{\'\i}sica Te\'orica, Facultad de CC. F{\'\i}sicas,
Univ. Complutense de Madrid, 28040 Madrid, Spain.\\
\small $^3$ 
Instituto de Ciencia de Materiales (CSIC). Cantoblanco,
28049 Madrid. Spain. \\
\small $^4$ Dip. di Fisica,
Univ. di Roma ``La Sapienza'',  00185 Roma and
INFN sezione di Roma, Italy.}
\date{\today}
\maketitle

\begin{abstract}
The Hybrid Monte Carlo algorithm is adapted to the simulation of a
system of classical degrees of freedom coupled to non self-interacting
lattices fermions. The diagonalization of the Hamiltonian matrix is
avoided by introducing a path-integral formulation of the problem, in
$d+1$ Euclidean space-time. A perfect action formulation allows to
work on the continuum euclidean time, without need for a
Trotter-Suzuki extrapolation. To demonstrate the feasibility of the
method we study the Double Exchange Model in three dimensions.  The
complexity of the algorithm grows only as the system volume, allowing
to simulate in lattices as large as $16^3$ on a personal computer.  We
conclude that the second order paramagnetic-ferromagnetic phase
transition of Double Exchange Materials close to half-filling belongs
to the Universality Class of the three-dimensional classical
Heisenberg model.
\bigskip

\noindent
05.10.Ln, 
75.10.-b, 
75.30.Et  
\end{abstract}

\section{Introduction}

Most of the models so far proposed to study the Colossal
Magnetoresistance manganites (CMR)~\cite{CMR}, share an extremely
simplifying feature: an assembly of non self-interacting lattice
fermion is coupled to an extensive number of classical continuous
degrees of freedom (the localized core spins of the Kondo model and of
the Double Exchange Model~\cite{DEM}, and/or the Jahn-Teller lattice
distortion fields~\cite{MILLIS}). Other physical context where this
simplifying feature appear are the pyrochlores or doubles perovskites.

The non self-interacting nature of the electrons in these models makes it
possible to explicitly perform the trace in Fock space, in terms of
the single-particle eigenstates. This yields a positive Boltzmann
weight for the continuous classical degrees of freedom, that for the sake
of brevity we will call spins in what follows (although they could be
a lattice distortion field!). In principle, the resulting problem
could be simulated by means of a Metropolis algorithm. However, the
update of a single spin requires a diagonalization of the
single-particle Hamiltonian matrix, which has a computational cost
proportional to the square of the lattice volume (if the most
sophisticated available algorithm is used). This implies that the time
needed to update all the spins on the lattice scales at best with the
cube of the lattice size. This problem has prevented the study of
systems with more than (say) two-hundred spins (a $6^3$ lattice) in
the simplest of the above quoted models, the Double Exchange model
(DEM), although most simulations~\cite{CALDERONBREY,DAGOTTO} are done
with a hundred or less spins. This is certainly not enough for an
accurate study of phase-transitions where most of the interesting
physics occurs.

In this paper, we reformulate the problem in the path-integral
formalism, obtaining an exact representation on $d+1$ dimensions for
the fermions and $d$ dimensions for the (classical) spins. In this
representation a positive Boltzmann weight is obtained, and the update
of the spins can be done by means of the Hybrid Monte Carlo (HMC)
algorithm~\cite{HMC}.  For the DEM, the computational cost of a
full-lattice updating is empirically found to grow as the lattice
volume (although a worst-case estimate would have yielded a
square-volume growing).  In addition, the autocorrelation time for HMC
is proportional to the correlation length while with the Metropolis
algorithm in the Hamiltonian formalism it grows like the correlation
length squared.  We will show that a standard simulation on a $4^3$
lattice yields fully compatible results with our HMC algorithm, but
the latter allows to simulate a $16^3$ lattice on a personal
computer. In this way we are able to obtain meaningful results for the
phase diagram of the DEM model. Some attention will be paid to the
largeness of the finite-size corrections on the small lattices.  We
will also show that in the absence of superexchange coupling between
the spins (whose numerical treatment is straightforward), the Double
Exchange Model near half-filling presents a second order phase
transition between the paramagnetic (PM) and the ferromagnetic (FM)
phase, that belongs to the Universality Class of the three-dimensional
Heisenberg model. Work is in progress for the study of the
phase-diagram of the DEM complemented with a first-neighbors
antiferromagnetic superexchange interaction. The final goal is to
confirm that the antiferromagnetic coupling is able to turn this PM-FM
phase transition from second to first order as predicted by
Mean-Field~\cite{DEMPH}. The phenomenological importance of reliably
finding PM-FM first-order transitions between phases of very similar
electronic densities cannot be overemphasized~\cite{UEHARA}.

The structure of the paper is as follows. In the next section we
describe the DEM, introducing our notational conventions and deriving
it from the Kondo lattice model. This somehow academic exercise will
allow to introduce in a natural way a
mathematically equivalent formulation of the DEM in terms of SU(2)
matrices rather than in terms of classical fixed-length spins (to this
respect, appendix~\ref{SU2} will be also of interest). 
This representation of the model will allow for an enormous improvement
of the numerical stability of the integration of the equations of motion
during the Molecular Dynamics part of the HMC algorithm.
In section 3, we present the path-integral formulation of
the model, and prove its mathematical equivalence with the Hamiltonian
one. In section 4, we give details of our implementation of HMC. Section 5
is devoted to consistency checks: we show
numerically how our {\em perfect action} formulation avoids the
need for a Trotter-Suzuki extrapolation to continuum Euclidean-time
and we compare the numerical results of the HMC simulation with an
usual Hamiltonian one.  
In section 6 we present our results for the
PM-FM phase transition at half filling. Section
7 is devoted to conclusions. We also include three appendices with
useful formulae and the proofs of some relations used.

\section{The model}

We consider the lattice Kondo model 
on a cubic lattice of
side $L$ and volume $V=L^3$, where periodic boundary conditions are
applied.  On each lattice site we have a classical localized spin,
\hbox{$\vec\phi_\vecx$} of unit-length. The spins interact with a band
of lattice fermions through the Hamiltonian
\begin{equation}
{\cal H}=\sum_{\vecx,\alpha}\sum_{\vecy,\beta}
c^{\dag}_{\vecx,\alpha} H_{\vecx,\alpha ; \vecy,\beta} c_{\vecy,\beta}\,,
\label{HAMILFOCK}
\end{equation}
where $\vecx$ and $\vecy$ run over all nodes of the spatial
lattice, and  \hbox{$\alpha,\beta= 1,2$}  
are spin indices. The single-particle Hamiltonian
matrix consist of a hopping term plus the Hund coupling with
the localized spins:
\begin{equation}
H_{\vecx,\alpha ; \vecy,\beta}=
-t\sum_{i=1}^d\delta_{\alpha,\beta}\left[ 
\delta_{\vecx; \vecy+\veci}+\delta_{\vecx; \vecy-\veci}\right] -J_H\, 
\delta_{\vecx; \vecy}(\vec\phi_\vecx\cdot\vec\sigma)_{\alpha,\beta}\, ,
\label{KONDO} 
\end{equation}
where $\veci$ is the unit vector in the $i$ direction and
$\vec\sigma=\{\sigma_1,\sigma_2,\sigma_3\}$ are the Pauli matrices.
For the particular case of the CMR manganites, the localized spins
represent the three core manganese t$_{2g}$ electrons that, due to the
Hund rule, yield a $S=3/2$ spin that for most purposes can
be considered as classical. The conduction electrons, represented by
the creation and annihilation operators $c_{\vecx,\alpha},\
c^{\dag}_{\vecx,\alpha}$, occupy the lowest of the two manganese e$_g$ 
orbitals, split by a Jahn-Teller distortion.

The statistical properties of the system with an explicit
superexchange antiferromagnetic coupling between the localized spins
can be obtained through the partition function. Choosing units such
that $k_{\mathrm B}=1$, the partition function reads
\begin{equation}
Z=\int \Dphi\  \e^{-\frac{1}{T}H^{SE}}\,\tr^{\mathrm{Fock}}\, 
\e^{-\frac{1}{T}(\cal H - \mu \cal N)}\, ,\label{PARTICIONFOCK}
\end{equation}
the superexchange hamiltonian being
\begin{equation}
H^{SE} =J_{\mathrm{AF}}\sum_{\vecx}\sum_{i=1}^d\, 
\vec\phi_\vecx\cdot\vec\phi_{\vecx+\veci}\,,
\end{equation}
and $\cal N$ is the number operator
\begin{equation}
{\cal N} = \sum_{\vecx,\alpha} c^{\dag}_{\vecx,\alpha}
c_{\vecx,\alpha}\, ,\quad [{\cal N},{\cal H}]=0\, .
\end{equation} 

The problem can be enormously simplified, due to the non
self-interacting nature of the Hamiltonian (\ref{HAMILFOCK}). Although
$\cal H$ is a $4^{V}\times 4^{V}$ matrix in the Fock space, the trace
in Eq.~(\ref{PARTICIONFOCK}) can be explicitly taken if the
eigenvalues of the $2V\times 2V$ single-particle Hamiltonian matrix,
$\{E_n\}_{n=1,\ldots,2V}$, are known:
\begin{equation}
\tr^{\mathrm{Fock}}\, \e^{-\frac{1}{T}(\cal H - \mu \cal
N)}=\e^{\sum_n\,\log{\left(1+\e^{-\frac{E_n-\mu}{T}}\right)}}\,.
\label{LOGEIGEN}
\end{equation}
It is thus clear that, as we have said in the introduction, the
resulting Boltzmann-weight is positive, and that the model can readily
be simulated by the Metropolis algorithm, up to the
computational caveats mentioned in the previous section. For numerical
calculations based on this strategy, see Ref.~\cite{DAGOTTO}.

The dimensionality of the matrices can be still reduced in a factor
of two, in the limit of large Hund coupling, thus obtaining Zener's
double-exchange model~\cite{DEM}. One first makes a unitary 
transformation that diagonalizes the Hund coupling term in 
Eq.~(\ref{KONDO}):
\begin{eqnarray}
H&\rightarrow & \Omega H \Omega^\dag\\
\Omega_{\vecx,\alpha;\vecy\beta}&=&\delta_{\vecx,\vecy}U({\vec\phi_\vecx})_{\alpha,\beta}\\[1mm]
U(\vec\phi)&=&\left(\begin{array}{rr}
\displaystyle \cos\frac{\theta}{2}\e^{\irm(\pi+\varphi)/2}&
\displaystyle \sin\frac{\theta}{2}\e^{\irm(\pi-\varphi)/2}\\[3mm]
\displaystyle \sin\frac{\theta}{2}\e^{\irm(\pi+\varphi)/2}&
\displaystyle -\cos\frac{\theta}{2}\e^{\irm(\pi-\varphi)/2}
\end{array}\right)\,,\label{Udef}
\end{eqnarray}
where $\theta$ and $\varphi$ are respectively the polar and azimuthal
angle that determine the spin $\vec\phi$ direction. It will also be
important in what follows our choosing of $U(\vec\phi)$ as an SU(2)
matrix. The resulting single-particle Hamiltonian matrix is
\begin{equation}
H_{\vecx,\alpha;\vecy,\beta}= -\ J_H\,(\sigma_3)_{\alpha,\beta}\ 
-t \sum_{i=1}^d\,\left[
\left(U(\vec\phi_\vecx)U^\dag(\vec\phi_{\vecy})\right)_{\alpha,\beta}
\delta_{\vecx,\vecy+\veci}\ +\ 
\left(U(\vec\phi_\vecx)U^\dag(\vec\phi_{\vecy})\right)_{\alpha,\beta}
\delta_{\vecx,\vecy-\veci}\,\right]\,. 
\label{HAMILGIRADA}
\end{equation}
Due to the largeness of the Hund coupling one should keep
only the electron state with spin parallel to its core spin (the ``1'' state
in the representation of Eq.~(\ref{HAMILGIRADA})). The
truncated single-particle Hamiltonian matrix is then
\begin{equation}
H_{\vecx,\vecy}= -t \sum_{i=1}^d\,\left[
\left(U(\vec\phi_\vecx)U^\dag(\vec\phi_{\vecx-\veci})\right)_{1,1}
\delta_{\vecx,\vecy+\veci}\ +\ 
\left(U(\vec\phi_\vecx)U^\dag(\vec\phi_{\vecx+\veci})\right)_{1,1}
\delta_{\vecx,\vecy-\veci}\right]\,.
\label{OURDEM}
\end{equation}
Let us take a look at the product
\begin{equation}
\left(U(\phi_\vecx)U^\dag(\phi_{\vecy})\right)_{1,1}=
\e^{\irm\varphi_\vecx/2}\left[
\cos\frac{\theta_\vecx}{2}\cos\frac{\theta_\vecy}{2}\ +\ 
\sin\frac{\theta_\vecx}{2}\sin\frac{\theta_\vecy}{2}\,
\e^{-\irm(\varphi_\vecx-\varphi_\vecy)}\right]
\e^{-\irm\varphi_\vecy/2}\, .
\label{LAEQSU2}
\end{equation}
The term between square brackets is nothing but the hopping term of
the DEM model (see e.g.~\cite{DEMPH}). Thus we see that the matrix in Eq.~(\ref{OURDEM}) is
actually an unitary-transformed of the usual hopping term, 
the unitary transformation
being
\begin{equation}
\tilde\Omega_{\vecx\vecy}=\delta_{\vecx,\vecy} \e^{\irm\varphi_\vecx/2}\,.
\end{equation}
Now, the expression
in Eq.~(\ref{OURDEM}) is extremely more convenient for an HMC study
than the usual one. In fact, during the Molecular Dynamics part of the
algorithm, one needs to take care of the constraint
$(\vec\phi_\vecx)^2=1$. It can be done with a modification of the
usual equation of motions as shown in Ref.~\cite{YUKAWA}.
To get these new equations of motion one needs to 
express the  hopping term of the DEM
in terms of the Cartesian coordinates of
the spin $(\phi^1,\phi^2,\phi^3)$
\begin{equation}
\cos\frac{\theta_\vecx}{2}\cos\frac{\theta_\vecy}{2}\ +\ 
\sin\frac{\theta_\vecx}{2}\sin\frac{\theta_\vecy}{2}\,
\e^{-\irm(\varphi_\vecx-\varphi_\vecy)}=
\frac{1}{2}\left(\sqrt{1+\phi^3_{\vecx\vphantom{y}}}\sqrt{1+\phi^3_\vecy}+
\frac{(\phi^1_\vecx-\irm \phi^2_\vecx)(\phi^1_\vecy+\irm \phi^2_\vecy)}
{\sqrt{1+\phi^3_{\vecx\vphantom{y}}}\sqrt{1+\phi^3_\vecy}}\right)\,.
\label{CARTESIAN}
\end{equation}
Indeed, a working HMC algorithm can be obtained using the above
representation~\cite{YUKAWA}, which is not analytic at the sphere
South Pole. However, during the Molecular Dynamics step of the HMC,
one needs the derivatives of the right-hand side of
Eq.~(\ref{CARTESIAN}), which at the South Pole are even more singular
than (\ref{CARTESIAN}), resulting on a poor numerical stability of the
integration of the equation of motion. On the contrary, the expression
of the hopping term as a function of the SU(2) matrices is
smooth. Moreover, as discussed in Appendix~\ref{SU2}, nothing changes
if we substitute the integrations over the spin-field in the partition
function, by an integration over the SU(2) group. If needed, the spins
$\vec\phi_\vecx$ can be obtained from the SU(2) matrices using the
formula (see appendix~\ref{SU2})
\begin{equation}
\phi_\vecx^j=\frac{1}{2}\tr\left(
\sigma_j U^\dag_\vecx \sigma_3 U_\vecx\right)\,,\ j=1,2,3\,.
\end{equation}
Thus we will consider the following
statistical system, which is strictly equivalent to Eq.~(\ref{PARTICIONFOCK}) 
in the double-exchange limit:
\begin{eqnarray}
Z&=&\int D U\e^{-\frac{1}{T}H^{SE}+\sum_{n=1}^V\,\log{\left(1+\e^{-\frac{E_n-\mu}{T}}\right)}}\, ,
\label{PARTICIONSU2}\\
H^{SE}&=& \frac{J_{\mathrm{AF}}}{2}\sum_{\vecx}\sum_{i=1}^d\, 
\tr\left[(U^\dag_\vecx\vec\sigma U_\vecx)\cdot
(U^\dag_{\vecx+\veci}\vec\sigma U_{\vecx+\veci})\right]\,.
\end{eqnarray}
In the above expression, $T$ is the temperature and $E_n$ are the
eigenvalues of the
single-particle Hamiltonian matrix defined in Eq.~(\ref{OURDEM}).

Although the SU(2) field $U_\vecx$ is still a constrained variable, it
can be dealt with using well established techniques from lattice-gauge 
theory~\cite{MONTVAYMUNSTER}. 

Let us also finally mention that the single-particle Hamiltonian
Eq.~(\ref{OURDEM}), is unitary equivalent to {\em minus} itself, the
unitary transformation being ($x,y,z$ are the lattice coordinates of
$\vecx$)
\begin{equation}
U_{\vecx,\vecy}=\delta_{\vecx,\vecy} (-1)^{x+y+z}
\label{HALF}
\end{equation}
This ensures that the spectrum is symmetric around zero and therefore
half-filling corresponds to $\mu=0$.
 
\section{From the Hamiltonian to the Path-Integral formulation}

In this section, we will show how to obtain a numerically tractable
path integral representation of the partition function
(\ref{PARTICIONSU2}) although our results will be valid for the
general problem outlined in the introduction:  classical continuous 
degrees of freedom coupled to non self-interacting fermions. 
In subsection~\ref{OBSFERMIONICOS} we shall also explain how some important
fermionic observables can be recovered in this formalism.

Let us first state the following  well known expression for the
partition function in terms of a pair of anticonmuting Grassman fields
$\{\Psi_\vecx^\dagger(\tau),\Psi_\vecx(\tau)\}$, where $\tau$ is the
Euclidean time~\cite{RMP},
\begin{eqnarray}
Z&=&\int D U \, D\Psi\, D\Psi^\dagger
\e^{-S_\mathrm{F}-\frac{1}{T}H^{SE}}\,,
\label{PARTICION}\\
S_F&=&\int_0^{\frac{\hbar}{T}}d\tau\left[ 
\sum_\vecx \Psi_\vecx^\dagger\frac{\partial \Psi_\vecx}{\partial \tau}
-\frac{1}{\hbar}\sum_{\vecx,\vecy}\Psi_\vecx^\dagger
(H_{\vecx,\vecy}-\mu\delta_{\vecx,\vecy})\Psi_\vecy\right] \label{COHERENTE}\,.
\end{eqnarray}
In the above expressions, $H$ is the single-particle matrix defined in
Eq.~(\ref{OURDEM}), and the Grassman fields verify antiperiodic
boundary conditions in the Euclidean-time direction
\begin{equation}
\Psi_\vecx^\dagger(0)=-\Psi_\vecx^\dagger(\hbar/T)\quad ,\quad
\Psi_\vecx(0)=-\Psi_\vecx(\hbar/T)\,.
\end{equation}

Now, in order to transform the representation (\ref{COHERENTE}) onto a
numerically tractable problem, we introduce a time discretization.  We
introduce $L_\tau$ time slices (for technical reasons, $L_\tau$ will
always be an even number, see Eq.~(\ref{POSITIVIDAD})), with a spacing 
$a_\tau$ such that
\begin{equation}
L_\tau a_\tau = \frac{\hbar}{T}\,.
\end{equation}
In this way, instead of a three-dimensional lattice and a continuum time,
we have a four dimensional lattice.
The Grassman fields now depend on the discrete coordinate $x_\tau$,
\begin{equation}
\tau= x_\tau a_\tau\quad,\quad x_\tau=0,1,\ldots,L_\tau-1\,,
\end{equation}
and verify the boundary conditions
\begin{equation}
\Psi_{0,\vecx}^\dagger=-\Psi_{L_\tau,\vecx}^\dagger\quad ,\quad
\Psi_{0,\vecx}=-\Psi_{L_\tau,\vecx}\,.\label{ANTISIMRET}
\end{equation}
The fields $U_\vecx$ instead, being classical, do not depend on
Euclidean time.  In order to check how close our time discretization
is from the continuous limit of Eq.~(\ref{COHERENTE}), we need to
compare $a_\tau$ with the natural time unit of our problem, $\hbar/t$ (see
Eq.~(\ref{OURDEM})). Therefore, the dimensionless parameter
that controls how close we are to the continuum-time limit is
\begin{equation}
\lambda=\frac{a_\tau t}{\hbar}\,.
\end{equation}
Our discretization should be such that in the $\lambda\to 0$ limit 
Eq.~(\ref{COHERENTE}) is recovered, much in the spirit of the Trotter-Suzuki
extrapolation.
From now on let us also adopt the convention that the quantities with
dimension of energy, $T,\mu,J_{\mathrm{AF}}$ and the matrix $H$ are measured in
units of $t$, in such a way that for instance
\begin{equation}
T=\frac{1}{\lambda L_\tau}\,.
\end{equation}
With all our notational conventions settled, the discretized form of
the action is
\begin{eqnarray}
S^\lambda_\mathrm{F}&=&\sum_{x_\tau,\vecx} 
\e^{\mu\lambda}\Psi^\dagger_{x_\tau,\vecx}
\Psi_{x_\tau+1,\vecx} -
\sum_{x_\tau,\vecx,\vecy}
\Psi_{x_\tau,\vecx}
\left[\exp(\lambda H)\right]_{\vecx; \vecy}
\Psi_{x_\tau,\vecy}\label{ACCIONDISCR}\\
&\equiv&\sum_{x_\tau,\vecx} \sum_{y_\tau,\vecy} \,
\Psi^\dagger_{x_\tau,\vecx} M^\lambda_{x_\tau,\vecx;y_\tau,\vecy}
\Psi_{y_\tau,\vecy}\,.\label{FERMIONICMATRIX}
\end{eqnarray}
The last equality on the above expressions defines the so called
fermionic matrix $M^\lambda$. The rationale for including the chemical
potential on the temporal {\em link} that joins the $(x_\tau,\vecx)$
site with the $(x_\tau+1,\vecx)$ one, can be found in
Refs.~\cite{POTQUIM}. It is easy to check that in the $\lambda\to 0$
limit the continuum-time action is recovered. The exponential form in
the spatial part of Eq.~(\ref{ACCIONDISCR}) is preferred over more
straightforward ones because it yields a {\em perfect} action, as
shown below, without any time discretization effect. For the
particular case of the DEM model, the action in
Eq.~(\ref{ACCIONDISCR}) can be directly simulated. For other models,
the approximated form
\begin{equation}
\left[\exp(\lambda H)\right]_{\vecx; \vecy}\approx \delta_{\vecx; \vecy}\ +\ 
\lambda\,H_{\vecx; \vecy}\,,
\label{TAYLORPRIMO}
\end{equation}
could be the only feasible one, but it would makes mandatory to
consider the $\lambda\to 0$ extrapolation.  

To show the correctness of our path-integral, it is useful to first
introduce the time Fourier transformed field,
\begin{equation}
\Psi_{x_\tau,\vecx}=\frac{1}{\sqrt{\Lt}}\sum_{p_\tau}\e^{i p_\tau
x_\tau}\Psi_{p_\tau,\vecx}\ ,\label{MATSUBARA} 
\end{equation}
where the sum extends over the Matsubara frequencies (see
Eq.~(\ref{ANTISIMRET})),
\begin{equation}
p_0=\frac{2\pi}{\Lt}q\,,\quad q=-\frac{\Lt-1}{2},\ldots ,-\frac{1}{2},
\frac{1}{2},\ldots,\frac{\Lt-1}{2}\,.\label{MATSUFRE}
\end{equation}
The fermionic action now reads
\begin{equation}
S^\lambda_\mathrm{F}=\sum_{p_0,\vecx} 
\e^{\mu\lambda+ip_0}\Psi^\dagger_{p_0,\vecx}\Psi_{p_0,\vecx} -
\sum_{p_0,\vecx,\vecy}  
\Psi^\dagger_{p_0,\vecx} \left[\exp(\lambda H)\right]_{\vecx;\vecy}
\Psi_{p_0,\vecy}\,.
\end{equation}
Therefore, the fermionic matrix defined in Eq.~(\ref{FERMIONICMATRIX})
is block-diagonal in this basis
\begin{eqnarray}
M^\lambda_{p_0,\vecx;p_0',\vecy}&=&\delta_{p_0,p_0'}
A_{\vecx;\vecy}(p_0,\lambda)\,,\label{BLOCK}\\
A_{\vecx;\vecy}(p_0,\lambda)&=&
\e^{\mu\lambda+ip_0}\delta_{\vecx,\vecy}\ -\  
\left[\exp(\lambda H)\right]_{\vecx;\vecy}\,,\label{LAAP0}
\end{eqnarray}
and the Hamiltonian matrix, being Hermitian, verifies
\begin{equation}
A^{\dag}(p_0,\lambda)=A(-p_0,\lambda)\, .
\end{equation}
The partition function is then (using the Grassman version of Gaussian 
integration)
\begin{eqnarray}
Z^\lambda&=&\int \prod_\vecx \d U_\vecx\,\prod_{q_0,\vecy} 
\d\Psi_{q_0,\vecy}\,
\prod_{p_0,\vecx} D\Psi^\dagger_{p_0,\vecx} \,
\e^{-\frac{H^{SE}}{T}-\sum_{p_0,\vecx}\sum_{q_0,\vecy}
\Psi^\dagger_{p_0,\vecx} M^\lambda_{p_0,\vecx;q_0,\vecy}\Psi_{q_0,\vecy}}\,, \\
&=& \int DU\,\det \left[M^\lambda\right]\, 
\e^{-\frac{1}{T}H^{SE}}\ =\ 
\int DU\,\prod_{p_0>0}\det \left[A^{\dag}(p_0,\lambda)A(p_0,\lambda)\right]\, 
\e^{-\frac{H^{SE}}{T}}
\,\label{POSITIVIDAD} .
\end{eqnarray}
It is then clear that the block-diagonal from of the fermionic matrix
yields a positive-definite Boltzmann weight. Now, in order to relate
this Boltzmann weight and our target expressed in
Eq.~(\ref{PARTICIONSU2}), let us first notice that the eigenvalues of
the $A(p_0)$, in terms of the $V$ eigenvalues of the Hamiltonian
single-particle matrix (\ref{OURDEM}), are
\begin{equation}
E_n^{A}(\lambda,p_0)=\e^{\mu\lambda+ip_0}-\e^{\lambda E_n}\,.
\end{equation}
Now using the equation
\begin{equation}
\prod_{p_0} \left(\e^{\mu\lambda+ip_0}-\e^{\lambda E}\right)=
\e^{\log{\left(1+\e^{-\frac{E-\mu}{T}}\right)+\frac{E}{T}}}\,,
\label{PRODUCTO}
\end{equation}
proved in appendix~\ref{IDENTIDADES},
we find for the fermionic determinant
\begin{eqnarray}
\det \left[M^\lambda\right]&=& \prod_n \prod_{p_0} \left(\e^{\mu\lambda+ip_0}-
\e^{\lambda E_n}\right)\, ,\\
&=&\e^{\sum_n \left[\log\left(1+\e^{-\frac{E_n-\mu}{T}}\right)
+\frac{E_n}{T}\right]}\,,\\
&=&\e^{\frac{1}{T}\tr H\,+\,\sum_n \left[\log\left(1+\e^{-\frac{E_n-\mu}{T}}\right)\right]}\label{EUREKA}\,.
\end{eqnarray}
Since the single-particle Hamiltonian matrix~(\ref{OURDEM}) is
traceless, it is clear that the discretized action exactly reproduces
the target Boltzmann-weight (\ref{PARTICIONSU2}) and thus it can be
rightly called a {\em perfect action\/}. In the general case, though,
one would have to take out by hand the $\tr H/T$ from the Boltzmann weight.

Before ending-up, let us say a few words about the (in principle) non
local matrix $\exp[\lambda H]$. In a model such as the DEM, where the
eigenvalues of the single-particle Hamiltonian matrix are within some
{\em a priori} known bounds, it can be numerically computed with a
polynomial expansion as described in appendix~\ref{PERFECTA}. 
In other cases, although the energy must be bounded from below if the
system is to be stable, an upper bound may not be available.
Then one will be enforced to use an approximation such as 
Eq.~(\ref{TAYLORPRIMO}). It would
anyway be convenient to add a multiple of the identity matrix to the
single-particle Hamiltonian, in order to have a positive
spectrum. From the above analysis, it follows
that with the approximation Eq.~(\ref{TAYLORPRIMO}), 
the simulation would be exact for the
$\lambda$ dependent single-particle Hamiltonian
\begin{equation}
H^\lambda=\frac{1}{\lambda}\log(1+\lambda H)\,.
\end{equation}
Therefore, there would be a deformation of the spectrum (as one finds using
the Trotter-Suzuki formula at finite time-slicing), that would disappear
on the $\lambda\to 0$ limit. More bothersome, there would be
an {\em empty-band} dynamical effect. Indeed, even in the $\mu\to -\infty$
limit, where fermions should not influence the classical degrees of 
freedom, the $\tr H^\lambda$ term of Eq.~(\ref{EUREKA}) would be 
present, and as one has
\begin{equation}
\frac{1}{\lambda}\log(1+\lambda H)=
H-\frac{\lambda}{2}H^2+\ldots\,,
\end{equation}
the spins would definitively feel this spurious interaction, even
if $\tr H=0$. One can completely cure this problem, by using the
Boltzmann-weight
\begin{equation}
Z=\int DU \prod_{p_0>0}\det \left[
\frac{A^{\dag}(p_0,\lambda,\mu)A(p_0,\lambda,\mu)}
{A^{\dag}(p_0,\lambda,\mu=-\infty)A(p_0,\lambda,\mu=-\infty)}
\right]\, 
\e^{-\frac{H^{SE}}{T}}\,,
\label{ZEMPTY}
\end{equation}
that can be simulated, using a straightforward modification of the
HMC algorithm explained in section~\ref{HMC}, because the matrices
$A(p_0,\mu)$ commute for all values of $p_0$, $\lambda$ and $\mu$.

\subsection{Fermionic Operators}\label{OBSFERMIONICOS}
\subsubsection{Charge density}
Let us call $|\vecx\rangle$ the state localized on the lattice site
$\vecx$, and $|n\rangle$ the eigenvector of the single-particle Hamiltonian
matrix (\ref{OURDEM}) corresponding to the eigenvalue $E_n$. The charge
density on site $\vecx$, for the given configuration of the spin-field is
\begin{equation}
\rho_{\vecx}= \sum_{n=1}^V |\langle n|\vecx \rangle|^2 
\frac{1}{\e^\frac{E_n-\mu}{T}+1}\,, 
\end{equation}
while the average charge-density on the lattice is
\begin{equation}
\rho\equiv  \frac{1}{V}\sum_{\vecx} \rho_{\vecx} =\frac{1}{V}\sum_n
\frac{1}{\e^\frac{E_n-\mu}{T}+1}\,.
\end{equation}
Now, using (see appendix \ref{IDENTIDADES})
\begin{equation}
\frac{1}{\Lt}\sum_{p_0}
\frac{e^{\mu\lambda+ip_0}}
{\e^{\mu\lambda+ip_0}-\e^{\lambda E}}=\frac{1}{1+\e^{\frac{E-\mu}{T}}}
\label{SUMA}
\end{equation}
we obtain
\begin{equation}
\rho_\vecx=\frac{1}{\Lt}\sum_{p_0}\e^{\lambda \mu+\irm p_0} 
\sum_n\frac{|\langle n|\vecx \rangle|^2 }{\e^{\lambda \mu+\irm p_0}
-\e^{\lambda E_n}}=
\frac{1}{\Lt}\sum_{p_0}\e^{\lambda \mu+\irm p_0} 
\left[A(p_0,\lambda)\right]^{-1}_{\vecx,\vecx}\,,
\end{equation}
while
\begin{equation}
\rho=\frac{1}{V}\frac{1}{\Lt}\sum_{p_0}\e^{\lambda \mu+\irm p_0} 
\tr\left[A(p_0,\lambda)\right]^{-1}\,.
\end{equation}
In practice, we make use of the equality between the thermal
average of $\rho_\vecx$ and the one of $\rho$, since 
most of computer-time during a HMC simulation is spent in
the inversion of the  matrices $A(p_0,\lambda)$, which is done row-by-row.
We therefore only calculate one row of the inverse matrix, and store
the corresponding value of $\rho_\vecx$. 

\subsubsection{Fermionic Energy}

In the Hamiltonian formalism the energy (per spin)
for a given configuration of the spin degrees of freedom, is
obtained from the logarithmic derivative with respect to the inverse
temperature, $\beta$,  of the partition function,
and has the form
\begin{equation}
e=\frac{1}{V}\sum_n
\frac{E_n-\mu}{\e^\frac{E_n-\mu}{T}+1}+\frac{1}{V} H^{SE}\,.
\end{equation}

Using again Eq.~(\ref{SUMA}), we can write the first term of the RHS of
the previous equation as
\begin{equation}
e_\mathrm F=\frac{1}{V\Lt}\sum_{p_0}\e^{\lambda\mu+\irm p_0}\sum_n
\frac{E_n-\mu}{\e^{\mu\lambda+\irm p_0}-e^{\lambda E_n}}\,,
\end{equation}
and thus
\begin{equation}
e_\mathrm F=\frac{1}{\Lt}\sum_{p_0}\e^{\lambda \mu+\irm p_0}
\frac{1}{V}
\tr\left[(H-\mu)A^{-1}(p_0,\lambda)\right]\,.
\end{equation}
As in the case of the density, one cannot afford to calculate
the full trace, but rely on the translational invariance and calculate
\begin{equation}
e_\mathrm F(\vecx)=\frac{1}{\Lt}\sum_{p_0}\e^{\lambda \mu+\irm p_0}
\left[(H-\mu)A^{-1}(p_0,\lambda)\right]_{\vecx,\vecx}\,,
\label{ELOCAL}
\end{equation}
that can be readily obtained once the $\vecx^\mathrm{th}$ row of the matrix
$A(p_0,\lambda)$  is known from the density calculation.

On the other hand, the total specific heat cannot be calculated in a practical 
way from the thermal
fluctuation of the energy. Indeed, one can easily find that
\begin{equation}
-\frac{\partial \langle e \rangle}{\partial \beta}=V\left(\left\langle
\frac{\partial e_F}{\partial \beta}
+e^2
\right\rangle - \left\langle e \right\rangle^2\right)\,.
\end{equation}
A representation analogous to Eq.~(\ref{ELOCAL}) can be readily
obtained for $\partial e_\mathrm{F}/\partial\beta$. The real problem
is the calculation of $\langle e^2\rangle$, because we do not know
$e_\mathrm{F}$, but $e_\mathrm{F}(\vecx)$. It is easy to convince oneself
that to substitute $e_\mathrm{F}$ by  $e_\mathrm{F}(\vecx)$ on the 
calculation of $\langle e^2\rangle$ produces a systematic overestimation,
magnified by the $V$ prefactor.

\section{Our implementation of HMC}\label{HMC}

In this section, we will give the necessary details about our implementation
of the HMC algorithm~\cite{HMC}. The reader interested in a full
exposition of the algorithm may consult~\cite{MONTVAYMUNSTER}.

Let us recall that we want to simulate the statistical
system
\begin{equation}
\int DU\,\prod_{p_0>0}\det \left[A^{\dag}(p_0,\lambda)A(p_0,\lambda)\right]\, 
\e^{-\frac{H^{SE}}{T}}\,.
\end{equation}
As usual, the first step is to get rid of the fermionic determinant by
using Gaussian integration, introducing the $\Lt/2$ pseudofermionic 
(commuting) fields, $\varphi_{p_0,\vecx}$:
\begin{equation}
\det M^\lambda=\int 
\left(\prod_{p_0>0,\vecx}\d \varphi_{p_0,\vecx}\right)
\left(\prod_{p_0>0,\vecx}\d \varphi_{p_0,\vecx}^*\right)
\exp\left\{-\sum_{p_0>0,\vecx,\vecy}\varphi_{p_0,\vecx}^*
(A^\dagger(p_0,\lambda) A(p_0,\lambda))^{-1}_{\vecx,\vecy}
\varphi_{p_0\vecy}\right\}\,.
\end{equation}
In our case of constrained variables belonging to the SU(2) group,
one introduces 3V {\em momenta} (one per group generator and per lattice
site~\cite{HMCSU2}),
by multiplying Eq.~(\ref{POSITIVIDAD}) by unity written in the form
\begin{equation}
1=\int_{-\infty}^{\infty}\,\prod_\vecx\,\frac{\d \vec P_\vecx}{(2\pi)^{3/2}}
\;\exp\left[-\frac{\vec P_\vecx^2}{2}\right]
\,.
\label{MOMENTA}
\end{equation}
So we end-up with a classical-mechanics model, that can be studied
using the Molecular Dynamics method. In this model the {\em kinetic} energy
is
\begin{equation}
{\cal T}=\sum_\vecx \frac{1}{2}\vec P^2_\vecx\,,
\end{equation}
while the {\em potential} one is
\begin{equation}
{\cal U}=\frac{H^{SE}}{T}+ \sum_{p_0>0,\vecx,\vecy}\varphi_{p_0,\vecx}^*
(A^\dagger(p_0,\lambda) A(p_0,\lambda))^{-1}_{\vecx,\vecy}
\varphi_{p_0\vecy}\,.
\end{equation}
Following the standard procedure, at the beginning of each Molecular
Dynamics trajectory, the momenta are extracted with the corresponding
Gaussian probability~(\ref{MOMENTA}), while the pseudofermions are obtained
from a Gaussian vector $\xi_{p_0,\vecx}$ as 
\begin{equation}
\varphi_{p_0}=A^\dagger(p_0,\lambda)\xi_{p_0}\,.
\end{equation}
In practice, the pseudofermions being instantaneously thermalized,
they are not changed during the trajectory. It is useful to consider
instead two molecular-dynamics time dependent fields
\begin{eqnarray}
\eta&=&(A^\dagger A)^{-1}\varphi\,,\label{LAETA}\\
\xi&=&A\eta\,,
\end{eqnarray}
Although $\varphi$ is not changing, the matrix $A(p_0,\lambda)$ changes
when the field $U_\vecx$ follows the dynamic. The equations of motion adapted
to the SU(2) group constraints are~\cite{HMCSU2} (the $\partial_{\vecx,j}$
derivative is defined in appendix~\ref{SU2}).
\begin{eqnarray}
\dot U_\vecx &=& \left(\irm \vec P_\vecx\cdot\sigb\right)U_\vecx\,,\\
\dot P_{\vecx,j}&=&-\partial_{\vecx,j} {\cal U}\,.
\label{EQMOVSU2}
\end{eqnarray}
The hard part to calculate is of course
\begin{equation}
\begin{array}{rcl}
\displaystyle
\partial_{\vecx,j} \left(\varphi^\dagger (A^\dagger
A)^{-1}\varphi\right)&=&\displaystyle-\varphi^\dagger (A^\dagger A)^{-1}
\left[(\partial_{\vecx,j}A^\dagger) A
+A^\dagger \partial_{\vecx,j} A\right]
(A^\dagger A)^{-1}\varphi\\\\
&=&\displaystyle-\eta^\dagger (\partial_{\vecx,j} A^\dagger)\xi -\xi (\partial_{\vecx,j} A)\eta=
-2\mathrm{Re}\left[ \xi^\dagger (\partial_{\vecx,j} A )\eta\right] 
\end{array}
\end{equation}
Thus we see that a knowledge of the full inverse $A(p_0,\lambda)$
matrices is useless, and it is enough to consider the field $\eta$
defined in Eq.~(\ref{LAETA}). Once we know how to calculate
derivatives of the exponential of the single-particle matrix (see
appendix~\ref{PERFECTA}), the rest of the calculation is standard: we
numerically integrate the equations of motion by means of the SU(2)
leap-frog algorithm~\cite{HMCSU2}, inverting the $A(p_0,\lambda)$
matrices using a conjugate-gradient method. A numerical trick of some
relevance is that one can calculate the inverses during the
Molecular-Dynamics steps of the algorithm with far less accuracy than
during the Monte Carlo accept-reject step~\cite{YUKAWA,GUPTA}. For the
exponential of the single-particle Hamiltonian, we have used an order
of the polynomial expansion such that the error is smaller than
$2\times 10^{-4}$ all over the spectrum.

An important remark about the algorithm is that pseudofermionic (four
dimensional) variables $\varphi$ can be straightforwardly generated
following the exact probability distribution. This allows to simulate
systems very long in the time direction without compromising the
autocorrelation time.

\section{Consistency checks}

We consider in this section some tests performed to check the
algorithm. Firstly, we should mention that although the computer code
for the HMC is rather complex, most of the routines are very easy
to check. For instance, the matrix inversion is self consistent, and
the integration of the equations of motion can be directly checked as
they should conserve the Molecular Dynamics Hamiltonian (up to second
order in the leap-frog step). In addition, we have checked explicitly
the reversibility of the equations of motion.

A posteriori, it is very useful to control the Creutz parameter\cite{CREUTZPAR}
defined as
\begin{equation}
\langle \e^{-\Delta H_\mathrm{MD}} \rangle\,,
\end{equation}
where all variations of $H_\mathrm{MD}$ must be considered (accepted
or not). This quantity should be 1, and its measure is a very strong
check of the simulation. A deviation would mean a reversibility
problem or a lack of equilibration. We have readily checked this
parameter in all the simulations.

\begin{figure}[t!]
\begin{center}
\epsfig{file=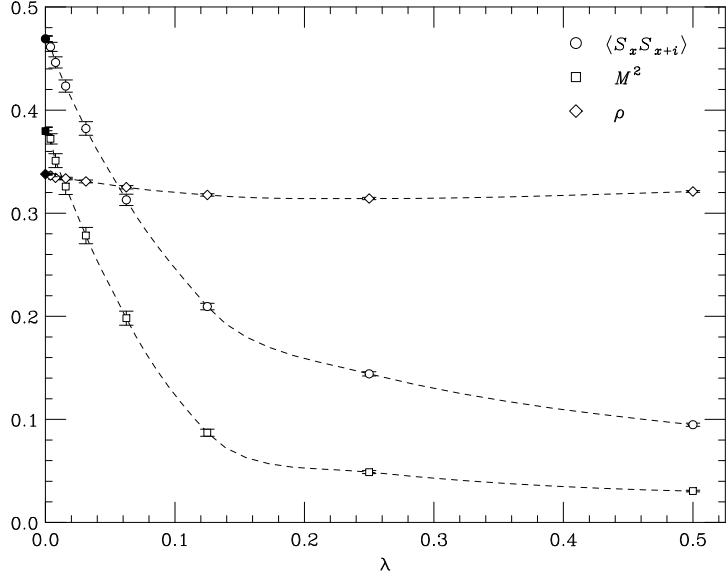,angle=90,width=0.6\linewidth}
\end{center}
\caption{Continuous time limit for the nearest neighbor correlation,
magnetization squared and charge density in a $4^3$ lattice
at $J_{\mathrm{AF}}/t=0$, $T=1/8$, $\mu=-0.35$. 
The left-most open symbols require $L_\tau=2048$. 
The filled symbols correspond to a
simulation with the Perfect Action ({\protect\ref{LEGEXP}}) using
$\lambda=0.25$, $L_\tau=32$.
\label{LIMITCONT}}
\end{figure}

Regarding the comparison of the time discretized model with the
physical continuous limit target, we have performed the following two
types of test. On the first place, we have simulated a $4^3$ lattice
at $T=1/8$, for decreasing values of $\lambda$, using
Eq.(\ref{TAYLORPRIMO}), with a shift of the identity $6\lambda$ that
ensures a positive spectrum. The empty-band dynamical effect is
avoided using Eq.(\ref{ZEMPTY}). We have chosen $\mu=-3.5$ and
$J_\mathrm{AF}=0$ which, for $T=1/8$ is near the
Paramagnetic-Ferromagnetic transition.  The results are displayed in
Fig. \ref{LIMITCONT} for several quantities. For this selection of the
parameters a linear behavior in $\lambda$ is observed only for large
values of $L_\tau=(T\lambda)^{-1}$.  We have also carried out a
simulation with the Perfect Action (see appendix \ref{PERFECTA}) in a
$4^3\times 16$ lattice at $\lambda=0.25$ with a 6 degree polynomial
approximation.  The result is plotted as filled symbols in
Fig. \ref{LIMITCONT}.  The agreement is excellent.  The selection of
$\lambda$ for a Perfect Action simulation should be taken looking at
the performance of the algorithm. Most of the results presented in
this article have been obtained with $\lambda=0.125$ and a polynomial
degree of 6.  Larger values of $\lambda$ have the advantage of
requiring smaller values of $L_\tau$ but the matrix inversion is more
expensive. Conversely, smaller values of $\lambda$ require larger
$L_\tau$ while the benefit in the matrix inversion is scarce.

Our second test, and maybe the strongest proof of the HMC method and
of our implementation of it is a direct comparison with numerical
results from a Hamiltonian simulation. The Hamiltonian model was
defined in terms of spins rather than SU(2) matrix, in order to
provide a full proof of equivalence. We have carried simulations with
both algorithms with the same parameters. We have chosen a $4^3$
lattice at $T=1/20$ for several values of the antiferromagnetic
coupling to go over the different phases of the system. Some of the
measures are presented in Table \ref{HMC-HAMIL}. We observe a perfect
agreement with precisions up to a few per thousand.

\begin{table}
\begin{center}
\begin{tabular*}{\hsize}{@{\extracolsep{\fill}}lllllll}
\hline\hline
$J_{\mathrm AF}/t$& 
&\multicolumn{1}{c}{$-0.01$}
&\multicolumn{1}{c}{0.05} 
&\multicolumn{1}{c}{0.2} 
&\multicolumn{1}{c}{0.3}\\
\hline
$\langle S_\vecx\cdot S_{\vecx+\veci} \rangle$
&Hamiltonian       & 0.7734(8) & 0.3817(18) & $-0.4699(5)$ & $-0.6842(5)$\\
&HMC               & 0.7717(10)& 0.3838(8)  & $-0.4697(3)$ & $-0.6852(4)$\\
\hline
$(\vec M)^2$
&Hamiltonian       & 0.7149(16)& 0.0162(7)  & $\hphantom{-}0.0130(4)$  
& $\hphantom{-}0.3580(13)$\\
&HMC               & 0.7127(16)& 0.0152(3)  & $\hphantom{-}0.01340(16)$
& $\hphantom{-}0.3611(11)$\\
\hline
\hline
\end{tabular*}
\end{center}
\caption{Comparison of the results of Hamiltonian and HMC simulations
in $L^3$ lattice with $T=1/20$ at half filling ($\mu=0$).
We show the correlation between nearest neighbor spins and the square
of the magnetization (magnetization staggered when that correlation is
negative). The numbers correspond to 10000 measures in each case.
\label{HMC-HAMIL}
}
\end{table}

\section{Numerical results.}

In this section we present the results of our HMC simulation using the
perfect action in the region of the Paramagnetic--Ferromagnetic phase
transition at vanishing superexchange coupling. We have chosen a fixed
temporal length $L_\tau=40$ varying the temperature through a $\lambda$
variation.

For simplicity on the analysis, we have restricted ourselves to the
half-filling case. Due to the hole-particle symmetry of the DEM, this
can be ensured by setting the chemical potential to zero (see
Eq.(\ref{HALF})). The study of other band-fillings requires to
carefully tune the chemical potential, and will be left for further
work.

\begin{figure}[ht]
\begin{center}
\epsfig{file=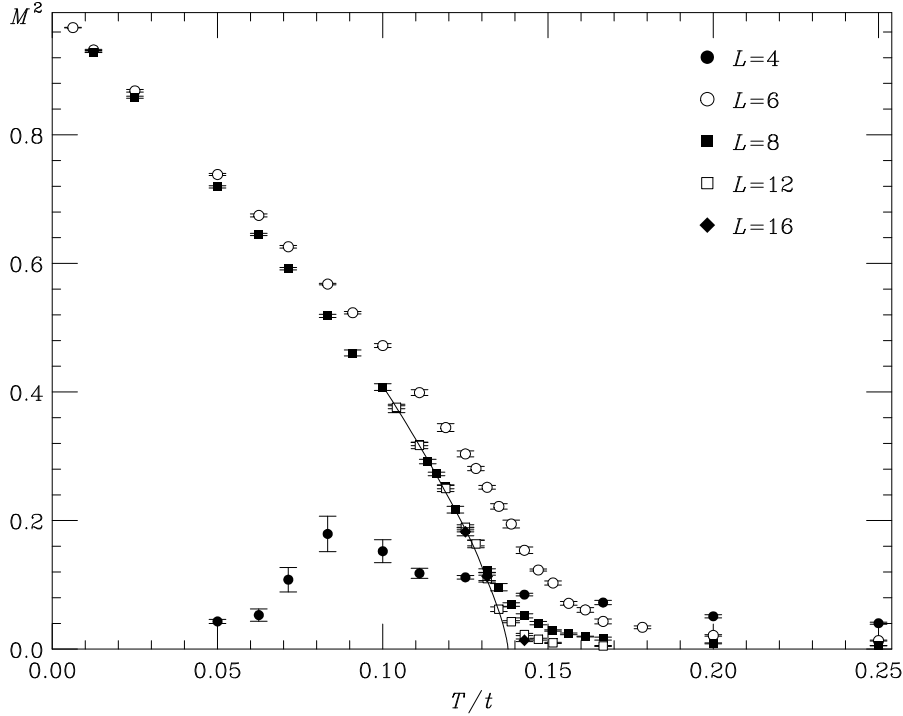,width=0.6\linewidth,angle=90}
\end{center}
\caption{Magnetization squared as a function of the temperature at
$J_\mathrm{AF}=0$, $\rho=0.5$ for several lattice sizes with
$L_\tau=40$.  The four (three) leftmost points for $L=6$($L=8$) have
been obtained with $\lambda=0.25$ and
$L_\tau=80,160,320,640$ ($L_\tau=80,160,320$).  The continuum line is a
fit to $A(T_\mathrm{c}-T)^{2\beta}$, with $\beta=0.37$, and
$T_{\mathrm{c}}$ taken from the $L=8,12$ lattices pair (see text).
\label{M2vTJ0}
}
\end{figure}

We have simulated in lattices of spatial sizes $L=4,6,8,12,16$ for
several values of the temperature. We measure every HMC trajectory
discarding up to 600 for thermalization (in the worst case). We
collect between 1000 and 10000 measures at every point.  We display
our results for the spin magnetization (squared) in
Fig.~\ref{M2vTJ0}. The time needed for a trajectory in a  500 MHz
Pentium III is about six minutes for a $12^3$ lattice in the critical
region with 25 leap frog steps of size 0.02.

Let us first define  the measured observables, and show their general
temperature and lattice-size evolution, and then later consider in detail
their behavior close to the critical region, and measure the critical
exponents.

The observables are best defined in terms of the correlation function
(the $\langle\cdot\rangle$ stands for Boltzmann average,
\begin{equation}
G(\vecx)=\frac{1}{V}\sum_\vecy 
\langle\vec\phi_\vecy\cdot\vec\phi_{\vecy+\vecx}\rangle\,,
\end{equation}
and its Fourier transform, $\hat G(\veck)$.
Then the susceptibility is proportional to the squared magnetization:
\begin{equation}
\chi=\hat G(\veck = 0) = V \langle M^2 \rangle
\end{equation}
It is also very useful to consider a finite-lattice correlation-length,
in terms of the minimum allowed momentum in a finite 
lattice~\cite{COOPER},
$\veck_{\mathrm{min}}=(2\pi/L,0,0)$
\begin{equation}
\xi=\frac{1}{2\sin (k_{\mathrm{min}})}
\left(\frac{\chi}
{\hat G(\veck_{\mathrm{min}})}-1\right)^{1/2}
\end{equation}
Notice that the above definitions use non-connected correlation
functions, and therefore the above correlation-length diverges in the
ferromagnetic phase like ${\cal O}(L^{1+D/2})$. In the thermodynamic
limit, $\xi$ diverges at the critical point like $|t|^{-\nu}$, ($t$ is
the reduced temperature).  The critical behavior for $\chi$ is
$|t|^{-\gamma}$.

In Fig.~\ref{M2vTJ0} we show the temperature and lattice size
evolution of $M^2$. There are several features to be noted.  The first
one is that the behavior of the $L=4$ lattice is rather pathological.
We believe that this evidences better than any other example the
need for larger lattices simulations of spin-fermion models. It is
also interesting to notice the larger lattices rapidly tend to their
thermodynamical limit, out from the critical region. Finally, we observe 
that the low temperature behavior of $M^2$ is linear. This can be
readily understood if we set that the average direction of the magnetization
is, say, the third axis. In that case,
\begin{equation}
M^2=1 - \frac{1}{V}\sum_\vecx\,\langle (\phi_\vecx^1)^2+
(\phi_\vecx^2)^2 \rangle+ {\cal O} ((\phi^1)^4,(\phi^2)^4,(\phi^1\phi^2)^2)\,.
\end{equation}
Since the deviations from the perfect ferromagnetic order are
proportional to the mean value of a quadratic operator, the linear
behavior with temperature follows from the equipartition principle,
that holds for our {\em classical} spins at low temperatures.

\begin{figure}[ht]
\begin{center}
\epsfig{file=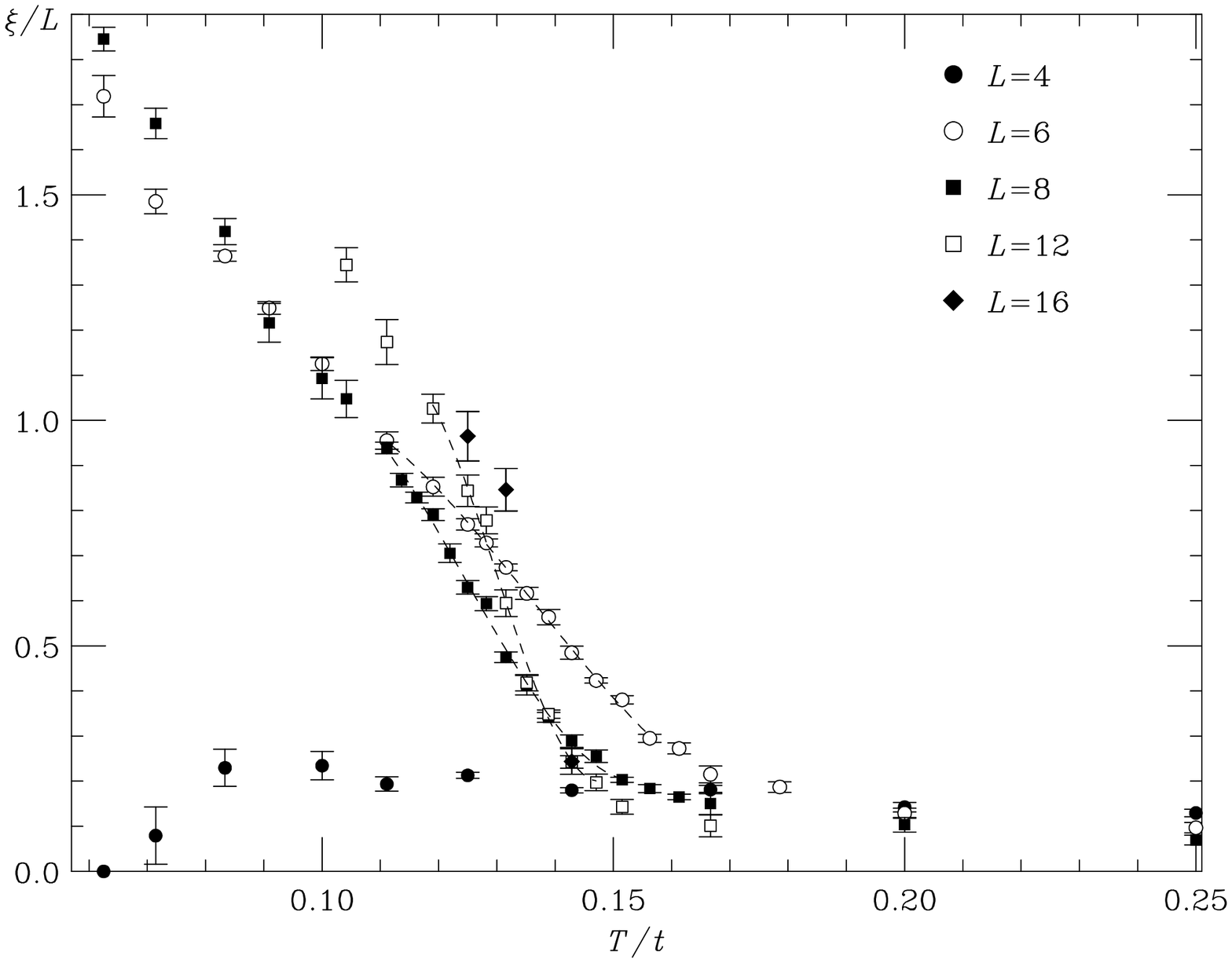,width=0.6\linewidth,angle=90}
\end{center}
\caption{Correlation-length in units of the lattice size as a function
of the temperature at $J_{AF}=0$, $\rho=0.5$ for several lattice sizes.
\label{xiLvT_J0}}
\end{figure}

In Fig.~\ref{xiLvT_J0} we show the correlation-length in units of
the lattice size. Notice that the curves for the different lattices
cross at a temperature growing with growing lattice-size. Eventually
the crossings should occur at the critical point, as dictated by
the Finite-Size Scaling Ansatz (see next section). One can also observe that
$\xi/L$ is a growing function of the lattice size in the ferromagnetic
phase, as it should be.

\subsection{Critical Exponents}

The main question of interest is whether the DEM presents a second
order phase transition between the paramagnetic phase and the ferromagnetic
one, at finite temperature. If the answer is positive, one may also wonder 
about the Universality Class of this phase transition~\cite{PARISI}. In 
principle, one of the two following scenarios should hold:
\begin{enumerate}
\item The ferromagnetic Double-Exchange interaction is long-ranged enough
to enforce Mean-Field behavior~\cite{MEANFIELD}. The critical exponents
would be $\nu=0.5$ and $\eta=0$.
\item The interaction is not long-ranged enough: the physical behavior 
should be the one of the classical Heisenberg model in three 
dimensions~\cite{PARISI}. The critical exponents would be $\nu=0.71(1)$ and
$\eta=0.041(2)$~\cite{O3}.
\end{enumerate}

In order to decide which of the above possibilities hold, we have applied 
the quotients-method~\cite{QUOTIENT}, to the Finite-Size Scaling 
Ansatz~\cite{FSSA}. We recall briefly the basis of this
method. Let $O$ be a quantity diverging in the thermodynamical limit
as $t^{-x_O}$ ($t=T/T_{\mathrm c}-1$ being the reduced
temperature). We can write the dependence of $O$ on $L$ and $t$ in the
following way~\cite{FSSA}
\begin{equation}
O(L,t)=L^{x_O/\nu} \left[F_O\left(\frac{L}{\xi(\infty,t)}  \right)+
{\cal O}(L^{-\omega},\xi^{-\omega}) \right] ,
\label{obs}
\end{equation}
where $F_O$ is a (smooth) scaling function and $(-\omega)$ is the
corrections-to-scaling exponent (e.g., $-\omega$ is the leading
irrelevant exponent of the Renormalization Group transformation). This
expression contains the not directly measurable term $\xi(\infty,t)$,
but if we have a good definition of the correlation length in a finite
box $\xi(L,t)$, Eq.~(\ref{obs}) can be transformed into
\begin{equation}
O(L,t)=L^{x_O/\nu} \left[G_O\left(\frac{\xi(L,t)}{L}  \right)+
{\cal O}(L^{-\omega}) \right] ,
\label{FSS2}
\end{equation}
where $G_O$ is a smooth function related with $F_O$ and $F_\xi$ and
the term $\xi_\infty^{-\omega}$ has been neglected because we are
simulating deep in the scaling region.  We consider the quotient of
measures taken in lattices $L$ and $sL$ at the same temperature
\begin{equation}
Q_O(s,L,t)=\frac{O(sL,t)}{O(L,t)}\,.
\end{equation}
Then, 
the main formula of the quotient method is 
\begin{equation}
\left.Q_O\right|_{Q_\xi=s}=
s^{x_O/\nu}+{\cal O}(L^{-\omega})\ ,
\label{QUO}
\end{equation}
i.e., we compute the reduced temperature $t$, at which the correlation
length verifies $\xi(sL,t)/\xi(L,t)=s$ and then the quotient between
$O(sL,t)$ and $O(L,t)$. In particular, we apply formula
(\ref{QUO}) to the overlap susceptibility, $\chi$, and the
$\beta$-derivative of the correlation length 
$\partial_T\xi$,
whose associated exponents are:
\begin{eqnarray}
x_{\partial_T\xi}&=&1+\nu\,,\\ 
x_{\chi}&=&(2-\eta)\nu .
\label{OBSERVABLES}
\end{eqnarray}
Notice that $\left.Q_O\right|_{Q_\xi=s}$ can be measured with great
accuracy because of the large statistical correlation between $Q_O$
and $Q_\xi$. It is also very important that in order to use
Eq.~(\ref{QUO}) one does not need the infinite-volume extrapolation for
the critical temperature.

In practice, what we do, is to perform a cubic polynomial fit to $\xi/L$
as a function of $T$, on the critical region and use the obtained continuous
function on the quotients formula (\ref{QUO}). We find
\begin{eqnarray}
\nu_{6,12}=0.75(4),&& T_{\mathrm{c}}=0.1284(9) t\\
\nu_{8,12}=0.72(9),&& T_{\mathrm{c}}=0.1379(6) t\,.
\end{eqnarray}
The above results are certainly compatible with the classical
Heisenberg model exponent, $0.71(1)$, and are 2.5 standard deviations
away from the Mean-Field result, $0.5$.  The estimate of the critical
temperature, shows a considerable lattice size dependency (it can be
shown that the crossing point tends to the critical point as
$L^{-1/\nu-\omega}$, $\omega$ being the universal scaling-corrections
critical exponent~\cite{QUOTIENT}). Using the crossing point for
(8,12) as an estimation of the critical temperature, we can perform a
fit of the magnetization squared to the function
$A(T_\mathrm{c}-T)^{2\beta}$. In Fig.~\ref{M2vTJ0} we show a fit with
the O(3) exponent $\beta=0.37$\cite{O3} (solid line). The MF value
would correspond to a linear behavior ($\beta=0.5$).  It seems
therefore safe to conclude that the second scenario is the one
realized in Double-Exchange materials with continous transitions,
which should have non MF critical beahaviour.  Let us however remark
that a really accurate measure of critical exponent would require the
extension of the reweighting techniques~\cite{FALCIONI} to these
models.

It is amusing to observe that the ratio between the real critical
temperature at half filling, $T_{\mathrm{c}}\approx0.14t$, and the
variational Mean Field estimate,
$T_{\mathrm{c}}^{\mathrm{MF}}=0.19t$~\cite{DEMPH}, is rather similar
to the corresponding ratio for the three dimensional classical
Heisenberg model ($T_{\mathrm{c}}=1.443J_{\mathrm{AF}}$~\cite{O3},
$T_{\mathrm{c}}^{\mathrm{MF}}=2 J_{\mathrm{AF}}$).
 
\begin{figure}
\begin{center}
\epsfig{file=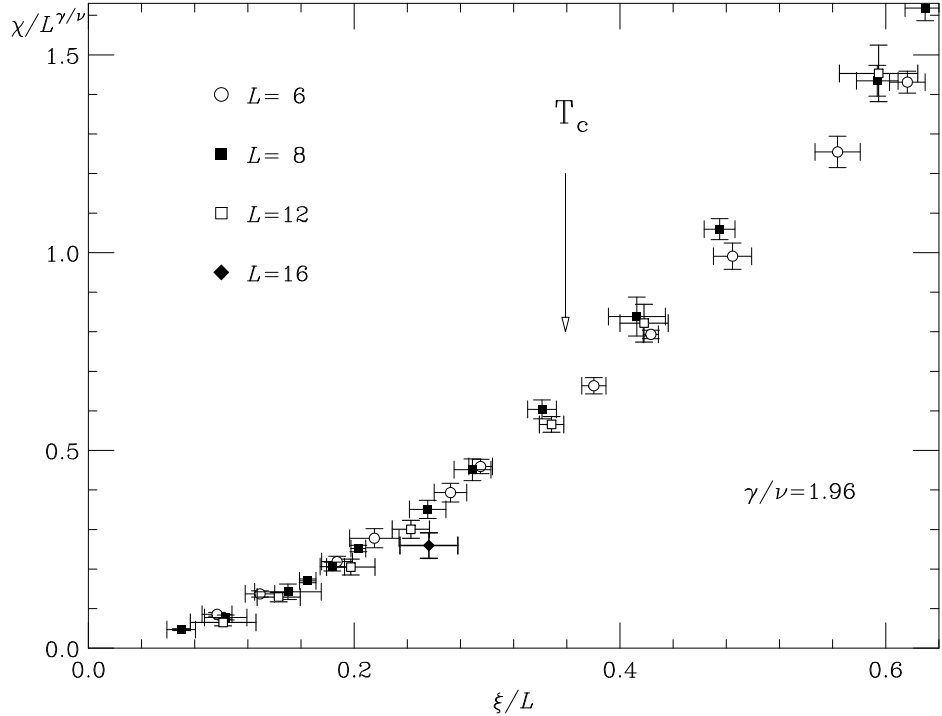,width=0.6\linewidth,angle=90}
\end{center}
\caption{Finite Scaling behavior of $\chi/L^{\gamma/\nu}$, as
a function of $\xi/L$, notice the data collapse. The arrow signals the
value of $\xi/L$ at the critical point. 
\label{FSS}}
\end{figure}

We finally perform the plot
suggested by Eq.~(\ref{FSS2}): $\chi/L^{\gamma/\nu}$ should be an
universal function of $\xi/L$. This seems to be rather well satisfied
by our data, with the critical exponents $\gamma$ and $\nu$ of the
classical Heisenberg model in three-dimensions.

\section{Conclusions and Outlook}

We have proposed a general numerical method for studying systems
consisting of classical degrees of freedom coupled to fermionic
fields. The method is based in the Path Integral formulation of
Quantum Mechanics that allows to work in a classical space-time
lattice where powerful Monte Carlo techniques, as the Hybrid Monte
Carlo method, are applicable since no sign problem arises.

As an example, we have describe explicitly the formulation of the
method in the case of the Double Exchange Model, observing that is
convenient to use a mapping of the spin to SU(2) matrices to avoid
singularities related with the parameterization of the Berry phase.

We have also shown that when the spectrum of the {\em single-particle}
Hamiltonian matrix is bounded, it is possible to work directly in the
continuum-time limit, using a {\em perfect action} thus avoiding the
need for a Trotter-Suzuki extrapolation. We have also shown how to
eliminate the spurious dynamical effects induced by the {\em empty}
fermion system, when the spectrum is unbounded.

We have finally presented some numerical results. First we have
described some consistency checks and then we have studied a property
of the model with direct physical interest as the
Paramagnetic--Ferromagnetic transition. We have studied the phase
transition at half-filling, where the transition temperature is
highest, for simplicity.  We have shown that the Finite Size Scaling
Ansatz is well satisfied for this model. The critical exponents have
turned out to be fully compatible with the ones of the three
dimensional classical Heisenberg model, and incompatible with the Mean-Field
prediction, as expected on Universality grounds if the interactions are
not extremely long-ranged. This conclusion was definitively out of reach
with the lattices that could be simulated with previous methods.

Work is in progress for the study of the full phase diagram of the
model, $(\rho,T,J_{\mathrm{AF}})$. We are also planning to use this Monte
Carlo method for the study of models with several electron orbitals
and/or phonons.

\section{Acknowledgements}
We acknowledge financial support from grants PB96-0875, AEN97-1680,
AEN97-1693, AEN99-0990 (MEC, Spain) and (07N/0045/98)
(C. Madrid). V.M.-M. is a M.E.C. fellow.  The simulations have been
carried out in RTNN computers at Zaragoza and Madrid.

\appendix

\section{Proof of Eqs.~(\protect\ref{SUMA}) and 
(\protect\ref{PRODUCTO})\label{IDENTIDADES}}

We recall that $1/T=L_\tau\lambda$ and that the sums (or products) in
$p_0$ run over the Matsubara frequencies (\ref{MATSUFRE}).

We apply the Poisson summation formula \cite{POISSON} (valid for a
continuous ($2\pi$)-periodic function)
\begin{equation}
\frac{1}{\Lt}\sum_{p_0}\ f(p_0)=
\sum_{s=-\infty}^{s=+\infty}\ (-1)^s \int_{-\pi}^{\pi}\frac{\d t}{2\pi}
\e^{\irm \Lt s t} f(t)\,
\end{equation}
to the RHS of Eq.~(\ref{SUMA}):
\begin{eqnarray}
\frac{1}{\Lt}\sum_{p_0} \frac{\e^{\irm p_0}}{\e^{\irm p_0}-\e^{\lambda
(E-\mu)}}&=& \sum_{s=-\infty}^{s=+\infty}\ (-1)^s
\int_{-\pi}^{\pi}\frac{\d t}{2\pi} \e^{\irm \Lt s t}
\frac{\e^{\irm t}}{\e^{\irm t}-\e^{\lambda (E-\mu)}}\, ,\\
&=&\sum_{s=-\infty}^{s=+\infty}\ (-1)^s \frac{1}{2\pi i}\int_{|z|=1}
\d z\ \frac{z^{s\,\Lt}}{z-\e^{\lambda (E-\mu)}}\,,
\end{eqnarray}
where the orientation of the contour is positive. For $s<0$ it is
useful to perform the integration in $w=1/z$.

When $\mu > E$, only the terms $s\geq 0$ contribute, and one obtains
\begin{equation}
\frac{1}{\Lt}\sum_{p_0} \frac{e^{\irm p_0}}{e^{\irm p_0}-\e^{\lambda
(E-\mu)}}=\sum_{s=0}^{\infty}\left[-\e^{\lambda
\Lt(E-\mu)}\right]^s
=\frac{1}{1+\e^{\frac{E-\mu}{T}}}\,,
\end{equation}
while if $\mu < E$, we need to consider only $s\leq -1$ arriving to
\begin{equation}
\frac{1}{\Lt}\sum_{p_0} \frac{\e^{\irm p_0}}{e^{\irm p_0}-\e^{\lambda
(E-\mu)}}=-\sum_{s=1}^{\infty}\left[-\e^{-\lambda
\Lt(E-\mu)}\right]^s\\
=\frac{1}{1+\e^{\frac{E-\mu}{T}}}\,.
\end{equation}

To prove the relation (\ref{PRODUCTO}), we start noting that 
(for $L_\tau$ even) the
products in its LHS can be grouped in pairs of nonzero complex
conjugates, so it is possible to write
\begin{equation}
\prod_{p_0} \left( \e^{\mu\lambda+\irm p_0}-\e^{\lambda E}\right)=
\e^{G(\mu,\lambda,E)}\,
\end{equation}
where the function $G(\mu,\lambda,E)$ is real. To obtain $G$ we 
first compute the $\mu$ derivative
\begin{eqnarray}
\frac{\partial G}{\partial \mu}&=& 
\sum_{p_0} \frac{\lambda \e^{\mu\lambda+\irm p_0}}
{\e^{\mu\lambda+ip_0}-\e^{\lambda E}}\\
&=&{\lambda \Lt} \frac{1}{1+\e^{\lambda \Lt(E-\mu)}}\\
&=&\frac{\partial}{\partial\mu}\log{\left(1+\e^{-\lambda \Lt(E-\mu)}\right)}\,.
\end{eqnarray}
From this relation, we know $G(\mu,\lambda,E)$ up to a
$\mu$-independent term $G_0(\lambda,E)$
\begin{equation}
G(\mu,\lambda,E)=
\log{\left(1+\e^{-\lambda \Lt(E-\mu)}\right)}+G_0(\lambda,E)\,.
\end{equation}
To evaluate $G_0$ it is enough to observe that
\begin{eqnarray}
\lim_{\mu\to -\infty} \e^{G(\mu,\lambda,E)} &=& (-1)^{\Lt} e^{\lambda \Lt E}\\
\lim_{\mu\to -\infty} \log{\left(1+\e^{-\lambda \Lt(E-\mu)}\right)}&=&0\,,
\end{eqnarray}
consequently, $G_0=E/T$ and Eq.~(\ref{PRODUCTO}) follows.

\section{Integrals over SU(2) and the sphere}\label{SU2}

In this appendix, we want to show that a generic integral over
the sphere
\begin{equation}
\int_{S^2} D \vec\phi f(\vec\phi) \equiv \frac{1}{4\pi}\int_{0}^{2\pi}
d\varphi\,\int_0^\pi\,\sin \theta\, d\theta\ f(\theta,\varphi)\,,
\end{equation}
can be substituted by an integral over the SU(2) group (with Haar's
invariant measure). 

In order to see how can this be possible, we start noticing that,
without loose of generality, the function depending on the vector
variable, $f(\vec\phi)$, can be considered as a function of the matrix
$(\vec\phi\cdot\vec\sigma)$, because
\begin{equation}
\phi_i=\frac{1}{2}\tr\,\left[\sigma_i(\phib\cdot\sigb)\right]\quad,\quad i=1,2,3\,.
\end{equation}
Now, one can always find an SU(2) matrix $U[\vec\phi]$, such that
\begin{equation}
U[\vec\phi] (\phib\cdot\sigb)\,U^\dag[\vec\phi]=\sigma_3\,.
\end{equation}
An explicit choice is given in Eq.~(\ref{Udef}).
There are two important facts to be noticed:
\begin{itemize}
\item Two SU(2) matrices, $V$ and $W$ verify $V^\dag \sigma_3 V=
W^\dag \sigma_3 W$ if, and only if, $V=\e^{\irm\alpha\sigma_3}W$ for
some $\alpha$, $-\pi<\alpha<\pi$.
\item For any SU(2) matrix, $W$, there is a point on the sphere $\vec\phi_W$,
such that $W^\dag \sigma_3 W = \vec\phi_W \cdot \vec\sigma$
\end{itemize}
Therefore, the SU(2) group can be parametrized as
\begin{equation}
W=\e^{\irm\alpha\sigma_3} U[\vec\phi]\,,\ -\pi<\alpha<\pi\,,\ \vec\phi^2=1\,. 
\end{equation}
The above considerations lead us to the following chain of equalities:
\begin{eqnarray}
\int_{S^2} \d \vec\phi f(\vec\phi)&=&\int_{S^2} 
\d \vec\phi f(\vec\phi\cdot\vec\sigma)\\
&=&
\int_{S^2} \d \vec\phi f\left(U^\dag[\vec\phi]\sigma_3 U[\vec\phi]\right)\\
&=&
\int_{S^2} \d \vec\phi\,\frac{1}{2\pi}\int_{-\pi}^{\pi}d\alpha\  
f\left(U^\dag[\vec\phi]\e^{-\irm\sigma_3\alpha}\sigma_3 \e^{\irm\sigma_3\alpha}U[\vec\phi]\right)\\
&=&\int_{SU(2)} \d W\, f(W^\dag\sigma_3 W)\,.
\end{eqnarray}
So we see that there is at least one integration measure over the SU(2) group,
for which our objective can be accomplished. The only thing that still remains
to be done is to show that the above integration measure is the proper 
Haar measure. It will be convenient to recall that the Haar measure is the
only one which is right invariant~\cite{CREUTZ}, namely for any function
$F$ over SU(2), and any SU(2) element $V$, one should have
\begin{equation}
\int_{SU(2)} \d W\,F(W)=\int_{SU(2)} \d W\,F(WV)\, .
\end{equation}
But, it is easy to see that if $W=\e^{\irm\sigma_3\alpha}U[\vec\phi]$, then
\begin{equation}
U[\vec\phi] V= \e^{\irm\beta(V,\vec\phi)\sigma_3} U[R_V \vec\phi]\,,
\end{equation}
where $R_V$ is the SO(3) rotation matrix associated with the SU(2) matrix
V, in the canonical homomorphism between both
groups~\cite{LIEGROUPS} 
\begin{equation}
[R_V \vec\phi]\cdot\vec\sigma = V^\dag (\vec\phi\cdot\vec\sigma)V\,.
\end{equation} 
At this point we can just go downhill:
\begin{eqnarray}
\int_{SU(2)} \d W\,F(WV)&=&
\int_{S^2} \d \vec\phi\,\frac{1}{2\pi}\int_{-\pi}^{\pi}d\alpha\ 
F\left(\e^{\irm(\beta(V,\vec\phi)+\alpha)\sigma_3} U[R_V \vec\phi] \right)\\
&=& \int_{S^2} \d \vec\phi\,\frac{1}{2\pi}\int_{-\pi}^{\pi}d\alpha\ 
F\left(\e^{\irm\alpha\sigma_3} U[R_V \vec\phi] \right)\\
&=& \int_{S^2} \d \vec\phi\,\frac{1}{2\pi}\int_{-\pi}^{\pi}d\alpha\ 
F\left(\e^{\irm\alpha\sigma_3} U[\vec\phi] \right)\,.
\end{eqnarray}
In the above expressions, the second equality follows from the periodicity
in $\alpha$ of the integrand, while the third is a consequence of the 
rotational invariance of the measure on the sphere. 

In order to formulate the Molecular Dynamics equations of motion,
one needs to know how to calculate derivatives on the SU(2) group.
For the shake of completeness, we give here the pertinent  definitions, 
but refer to \cite{MONTVAYMUNSTER} for a complete exposition.

One defines three different derivatives over SU(2) (one per group generator)
\begin{equation}
\partial_j f(U)=\left.\frac{\d\,f(\e^{\irm\epsilon\sigma_j}U)}
{\d\,\epsilon}\right|_{\epsilon=0}\,.
\end{equation}
If $f$ is a smooth function of the matrix element of $U$,
$U_{\alpha,\beta}$, we have
\begin{equation}
\partial_j f(U) = \sum_{\alpha,\beta} \frac{\partial f(U)}
{\partial U_{\alpha,\beta}}\left(\irm\sigma_j U\right)_{\alpha,\beta}\,.
\end{equation}
If it depends in the full lattice configuration, $\{U_\vecx\}$, we
define
\begin{equation}
\partial_{\vecx,j} f(U) = \sum_{\alpha,\beta} \frac{\partial f(U)}
{\partial (U_\vecx)_{\alpha,\beta}}\left(\irm\sigma_j
U_\vecx\right)_{\alpha,\beta}\,.
\end{equation}

\section{The exponential of the single-particle matrix}\label{PERFECTA}

In this appendix, we show how to numerically deal with the exponential
of a matrix, like the Double-Exchange single-particle Hamiltonian matrix,
with eigenvalues verifying $-6\le E_n\le 6$. 

Let us call $c_n^\lambda$ the coefficients of the Legendre
polynomials expansion of the function $e^{6\lambda x}$ for $x\in[-1,1]$. 
We can write
\begin{equation}
\e^{\lambda H_\mathrm{DEM}}=
\sum_{n=0}^{\infty}\, c_n^\lambda P_n(H_\mathrm{DEM}/6)\,.\label{LEGEXP}
\end{equation}
In the following, we shall use the shortcut $\hat H=H_\mathrm{DEM}/6$.
In practice we use the truncation 
\begin{equation}
Q(N,\lambda)=\sum_{n=0}^{N}\, c_n^\lambda P_n(\hat H)\,,
\end{equation}
that correspond to a Hamiltonian
\begin{equation}
H^T=\frac{\log Q(N,\lambda)}{\lambda}\,.
\end{equation}
The truncation error is quantified through the function
\begin{equation}
R_N(x,\lambda)=\frac{\log\left[\sum_{n=0}^{N}\, c_n^\lambda P_n\left(\frac{x}{6}\right)\right]}{\lambda}-x\quad,\quad x\in[-6,6]\,.
\end{equation}
that would be zero if the real exponential was calculated.
For instance, $R_{10}(x,1/2)<2\times 10^{-4}$ for all the interval.

To preserve the numerical stability is better to use the recurrence-relations 
of the Legendre polynomials than their actual expressions in terms of 
$\hat H$. Starting from
\begin{equation}
P_0(\hat H)=1,\quad P_1(\hat H)=\hat H\,,
\end{equation}
we will use (for $n>1$)
\begin{equation}
P_{n+1}(\hat H)|v\rangle =\frac{2n+1}{n+1} \hat H P_n(\hat H)|v\rangle - 
\frac{n}{n+1} P_{n-1}(\hat H) |v\rangle\,.\label{RECURRENCE}
\end{equation}
Notice that since matrix $\hat H$ is sparse (6 non-vanishing matrix 
element per row), the truncated expression for the exponential can be
calculated in order $V$ operations.

In the HMC, to integrate the equations of motion, we need to know the 
matrix elements
\begin{equation}
\langle G|\sum_{n=0}^N c_n^\lambda 
\frac{\delta P_n(\hat H)}{\delta U_\vecx}\,|F\rangle\,.
\end{equation}
From (\ref{RECURRENCE}) we can write a recursive relation for the
derivative. However it would mean a recursion (involving $O(V)$
multiplications) for each lattice site. This would make a total of $O(V^2)$
operations. 

Fortunately, it is possible to obtain the matrix elements
with $O(V)$ operations. To this end we use the double expansion
\begin{equation}
\frac{\delta P_n(\hat H)}{\delta U_\vecx}=
\sum_{m_1=0}^{n-1}\sum_{m_2=0}^{n-1-m_1}\, L^{(n)}_{m_1,m_2}\,
P_{m_1}(\hat H)\frac{\delta \hat H}{\delta U_\vecx} P_{m_2}(\hat H)\,.
\label{RECUD}
\end{equation}
In this equation $L^{(n)}_{m_1,m_2}$ are symmetric in $m_1,m_2$ and
vanish for $m_1+m_2\ge n$. They can be obtained from the following relations
\begin{itemize}
\item If $m_1+m_2\le n-2$
\begin{equation}\
L^{(n+1)}_{m_1,m_2}=\frac{2n+1}{n+1}\left[
\frac{m_1}{2m_1-1}L^{(n)}_{m_1-1,m_2}+
\frac{m_1+1}{2m_1+3}L^{(n)}_{m_1+1,m_2}\right]-
\frac{n}{n+1}L^{(n-1)}_{m_1,m_2}\,,
\end{equation}
\item if $n-1\le m_1+m_2\le n$, with $m_1\ne 0$
\begin{equation}
L^{(n+1)}_{m_1,m_2}=\frac{2n+1}{n+1}
\frac{m_1}{2m_1-1}L^{(n)}_{m_1-1,m_2}\,,
\end{equation}
\item Finally
\begin{equation}
L^{(n+1)}_{0,n}=\frac{2n+1}{n+1}\,.
\end{equation}
\end{itemize}

In terms of the $L$ coefficients we can write
\begin{equation}
\langle G|\sum_{n=0}^N c_n^\lambda 
\frac{\delta P_n(\hat H)}{\delta U_\vecx}\,|F\rangle=
\left(\sum_{m_1=0}^{N-1}\langle G|P_{m_1} (\hat H)\right)
\frac{\delta\hat H}{\delta U_\vecx} \left(
\sum_{n=0}^N\,c_n^\lambda\,\sum_{m_2=0}^{n-1-m_1}\,L^{(n)}_{m_1,m_2}\,
P_{m_2} (\hat H) |F\rangle\right)\,. 
\end{equation}

\end{document}